\documentclass[aps,prl,twocolumn,superscriptaddress]{revtex4-2}
\usepackage{graphicx,xcolor}
\usepackage{amsmath,amssymb,amsthm}
\usepackage{xr-hyper}
\usepackage[pdfencoding=auto]{hyperref}
\usepackage{physics}
\usepackage[small,bf]{caption}
\usepackage[subrefformat=parens, labelformat=simple, singlelinecheck=false]{subcaption}
\usepackage{filecontents}

\hypersetup{colorlinks=true,allcolors=black}
\theoremstyle{definition}
\newtheorem{theorem}{Theorem}

\newtheorem{corollary}[theorem]{Corollary}

\newtheorem{definition}[theorem]{Definition}

\begin{document}
\title{Stabilizer formalism in linear optics and application to Bell-state discrimination}
\author{Tomohiro Yamazaki}
\email{yamazaki-t@qi.mp.es.osaka-u.ac.jp}
\affiliation{Graduate School of Engineering Science, Osaka University, Toyonaka, Osaka 560-8531, Japan}
\affiliation{Center for Quantum Information and Quantum Biology, Osaka University, Toyonaka, Osaka 560-0043, Japan}

\author{Rikizo Ikuta}
\affiliation{Graduate School of Engineering Science, Osaka University, Toyonaka, Osaka 560-8531, Japan}
\affiliation{Center for Quantum Information and Quantum Biology, Osaka University, Toyonaka, Osaka 560-0043, Japan}

\author{Takashi Yamamoto}
\affiliation{Graduate School of Engineering Science, Osaka University, Toyonaka, Osaka 560-8531, Japan}
\affiliation{Center for Quantum Information and Quantum Biology, Osaka University, Toyonaka, Osaka 560-0043, Japan}

% \date{\today}
\begin{abstract}
    We propose a framework to analyze linear optical circuits based on an analogy with stabilizer formalism in quantum circuits, which provides efficiently computable formulas related to state discriminations.
    Hence, we analyze a Bell-state discrimination scheme with linear optics and ancillary single photons.
    With an increasing number of ancilla photons, the success probability of Bell-state discrimination has a maximum of $\frac{403}{512} \simeq 0.787$ at $28$ ancilla photons.
    By contrast, the corresponding two-qubit measurement asymptotically approaches a maximally entangling measurement.  
\end{abstract}
\maketitle

\textit{Introduction.---}
Linear optical quantum computation (LOQC) is a promising model of quantum computation consisting of photon generators, photon detectors, and passive linear optics.
Despite the different structure from qubit-based quantum computation,
it allows universal quantum computation by combining feed-forward operations~\cite{Knill2001-eq}.
While most of the underlying technology is currently maturing~\cite{Carolan2015-ou,Bogaerts2020-yq,Kim2020-vp,Gyger2021-ub}, the large-scale implementations remain challenging because of high resource overheads~\cite{Li2015-sq} in addition to experimental imperfections.
By regarding LOQC as computation under constraints of restricted feasible operations,
the protocols have been improved to measurement-based~\cite{Nielsen2004-bw, Browne2005-ft}, percolation-based~\cite{Kieling2007-ku, Gimeno-Segovia2015-dt}, and fusion-based protocols~\cite{Bartolucci2021-pc}.

The operations feasible with linear optical circuits (LOCs)~\footnote{Herein, LOCs generally refer to setups consisting of linear optical elements, photon-number resolving detectors, ancilla photons, and feed-forward operations. Depending on the context, it also refer to only setups consisting of linear optical elements.} themselves also need to be studied because their feasibility and success probability strongly affect the overall performance of LOQC.
Many schemes for new types of operations~\cite{Fiurasek2006-px,Cable2007-am,Tashima2009-mn,Zhang2019-dd,Luo2019-hx,Paesani2021-qn} and for increasing the success probabilities of desired operations~\cite{Grice2011-hn,Zaidi2013-db,Ewert2014-uw,Olivo2018-qj,Bartolucci2021-hk} have been found.
However, for brute-force explorations of such schemes, LOC outputs need to be calculated for every possible input state.
This task is computationally difficult because it is related to computing matrix permanents, which is a $\#$P-hard problem~\cite{Aaronson2011-nd}.
Therefore, finding a class of LOCs easily analyzable but still valuable for quantum computation is worthwhile.
Stabilizer formalism in quantum circuits, which we refer to as quantum stabilizer formalism (QSF), leads to the Gottesman-Knill theorem~\cite{Gottesman1998-vw}.
According to this, Clifford circuits with state preparations and measurements in the computational basis can be simulated efficiently.
Thus, this theorem establishes a class of easily analyzable quantum circuits.

Bell-state discriminations in dual-rail encoding are of primary importance.
Deterministic Bell-state discrimination is impossible~\cite{Vaidman1999-ad, Lutkenhaus1999-sy}, and the maximum success probability is 50\% without ancilla photons~\cite{Calsamiglia2001-zr,Weinfurter1994-kd, Braunstein1995-kz}.
Some schemes realize near-deterministic Bell-state discrimination but require larger entangled states as ancillae to yield higher success probability~\cite{Grice2011-hn,Ewert2014-uw}.
Thus, scalable schemes with a fixed size of entangled ancillae are practically important.

Here, we introduce bosonic stabilizer formalism (BSF), a framework similar to QSF but defined in LOCs.
Although BSF does not enable efficient simulations of all LOC outputs, the BSF-based classification simplifies their analysis.
The conditions for destructive interferences, known as suppression or zero-transmission laws, have been studied for various LOCs~\cite{Lim2005-wt,Tichy2010-lj,Tichy2012-ds,Crespi2015-op, Weimann2016-hu,Crespi2016-pi,Dittel2017-cl,Su2017-vn,Viggianiello2018-mz}.
We derive the general formula equivalent to the one derived in Ref.~\cite{Dittel2018-sz,Dittel2018-uk}.
In addition, we discuss further generalizations and a procedure to analyze an LOC as a quantum operation.

Additionally, we consider a scheme of Bell-state discrimination with ancillary single photons as a first sample application of BSF.
The success probability of Bell-state discrimination of up to $\frac{403}{512} \simeq 0.787$ is achieved; to the best of our knowledge, this is the highest value among the known Bell-state discrimination schemes without entangled ancillae.
The failure events also correspond to projections onto some entangled states.
When the number of ancilla photons increases, almost all those states approach the Bell states; thus, this scheme asymptotically approaches a maximally entangling measurement.
All proofs are presented in the Supplemental Material.

\textit{Bosonic stabilizer formalism.---}
We consider the evolutions of Fock states by $m$-port LOCs.
The transformations of the creation operators are expressed using a transfer matrix $U$ as $a_i^\dag = \sum_{j=0}^{m-1} a_j^\dag [U]_{ji}$ for $i=0,\cdots,m-1$, where $[U]_{ji}$ is the $(j, i)$ entry of $U$.
Any unitary matrix is realized as a transfer matrix~\cite{Reck1994-gm}; i.e., the set of all transfer matrices of $m$-port LOCs is $\mathrm{U}(m)$.
\begin{definition}[$n$-boson representation]\label{boson_rep}
    For an $m\times m$ matrix $S$ and $m$-tuples $\vb*{n}$ and $\vb*{n'}$, let
    \begin{equation}
        S^{(\vb*{n},\vb*{n'})} = \mqty([S]_{0 0} J_{n_0 n'_0} & \dots & [S]_{0 m-1} J_{n_0 n'_{m-1}} \\ \vdots &&\vdots \\ 
        [S]_{m-1 0} J_{n_{m-1} n'_0} & \dots & [S]_{m-1 m-1} J_{n_{m-1} n'_{m-1}} 
        ),
    \end{equation}
    where $J_{n_i n'_j}$ is the $n_i \times n'_j$ matrix, all elements of which are $1$, for integers $n_i, n_j$.
    Then, the $n$-boson representation $B_n$ is a group homomorphism from $\mathrm{U}(m)$ to $\mathrm{U}(N)$ such that 
    \begin{equation}
        [B_n(S)]_{\vb*{n},\vb*{n'}}=\frac{\text{Per}(S^{(\vb*{n}\vb*{n'})})}{\sqrt{\prod_{i=0}^{m-1} n_i! n'_i !}},
    \end{equation}
    for any $S\in \mathrm{U}(m)$ and m-tuples $\vb*{n}$ and $\vb*{n'}$ satisfying $\sum_i n_i =\sum_i n'_i =n$. Here, $N = \frac{(m+n-1)!}{(m-1)! n!}$,
    and $\text{Per}(S^{(\vb*{n},\vb*{n'})})$ is the matrix permanent of $S^{(\vb*{n},\vb*{n'})}$.
\end{definition}
\begin{figure}[tbp]
    \begin{minipage}[b]{0.43\linewidth}
        \centering
        \subcaption{}
        \includegraphics[scale=0.45, trim=0 0 0 0, clip]{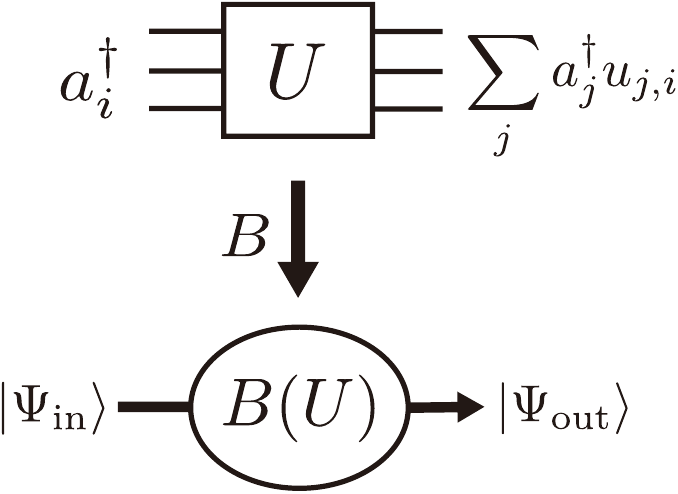}
        \label{image_boson_rep}
    \end{minipage}
    \begin{minipage}[b]{0.55\linewidth}
        \centering
        \subcaption{}
        \includegraphics[scale=0.5, trim=0 0 0 0, clip]{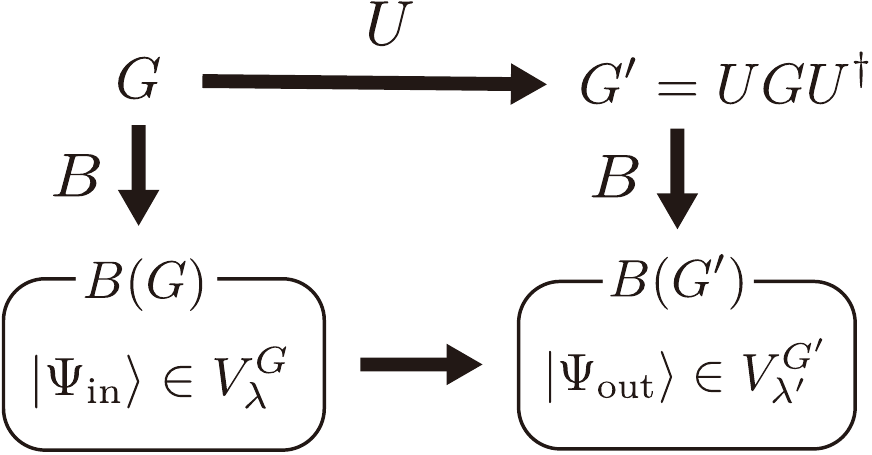}
        \label{image_BSF}
    \end{minipage}
    \caption{Conceptual diagrams of (a) $n$-boson representation and (b) BSF.}
    \label{image}
\end{figure}
The number of photons $n$ does not change until detection.
Thus, we consider only fixed-$n$ input states in the following.
Here, $B_n$ in Def.~\ref{boson_rep} is equivalent to the map from transfer matrices to the matrix representations of the corresponding unitary evolutions between the $n$-photon Fock states~\cite{Scheel2004-xy}.
For convenience, we define the map to the unitary evolution itself as $B$; i.e., its matrix representation is the block diagonal matrix $\bigoplus B_n(S)$ for $S\in \mathrm{U}(m)$.
Figure~\ref{image_boson_rep} is a conceptual diagram of Def.~\ref{boson_rep}. 

Computing $B_n(S)$ for $S$ is generally intractable, as 
the matrix permanent must be calculated for each element, and the matrix size $N$ exponentially increases with $n$ and $m$.
However, computing $B_n(S)$ is easy when $S$ is monomial, i.e., a product of a permutation matrix and a diagonal matrix. 
That is obvious because an LOC represented by a unitary monomial transfer matrix corresponds to independent phase shiftings followed by mode permutations. 
For simplicity, we introduce the following notations.
\begin{definition}
The $m \times m$ unitary monomial group $\mathcal{M}_m$ is the group of all $m \times m$ unitary and monomial matrices.
$D$ and $\sigma$ are maps from $\mathcal{M}_m$ to the set of all $m\times m$ diagonal matrices and the symmetry group $S_m$, respectively, satisfying $g=P_{\sigma_g}D(g)$ for $g\in \mathcal{M}_m$, where $P_{\sigma_g}$ is the permutation matrix corresponding to permutation $\sigma_g$.
\end{definition}
In QSF, a stabilizer group $G$ directly stabilizes a state $\ket*{\Psi}$ as $g\ket*{\Psi}=\ket*{\Psi}$ for any $g \in G$.
The main concept of BSF is to stabilize a Fock state using $B_n(G)$ instead of $G$ itself.
In BSF, an Abelian group $G$ is called a stabilizer group, and all joint eigenspaces of $B_n(G)$ are defined (not only that with $+1$ eigenvalues) as its stabilizer spaces.
\begin{definition}
    For an Abelian group $G$, $V^G_\lambda$ is a joint eigenspace of $B_n(G)$ with eigenvalues represented by $\{\lambda(g)|g\in G\}$, where $\lambda$ is a function. Here, $\rho_\lambda^G$ is the projector onto $V^G_\lambda$.
\end{definition}

By taking $G\subset \mathcal{M}_m$ as a stabilizer group, $B(G) \subset \mathcal{M}_N$ is easily calculable.
The following theorem addresses the three basic considerations for BSF: how to (i) calculate stabilizer spaces, (ii) handle state evolutions, and (iii) be related to measurement outcomes in the Fock basis.
\begin{theorem}\label{main_thm}

(i) For an Abelian group $G\subset \mathcal{M}_m$ and a state $\ket*{\Psi} = \sum_{\vb*{n}} c_{\vb*{n}} \ket*{\vb*{n}}$,
\begin{equation}\label{projection_formula}
    \norm{\rho_\lambda^G \ket*{\Psi}}^2 = \frac{1}{\abs{G}}\sum_{g\in G} \lambda(g)^{-1}\sum_{\vb*{n}}c_{\sigma^{-1}_g(\vb*{n})}^* c_{\vb*{n}}\prod_{i=0}^{m-1} [D(g)]_{i i}^{n_i},
\end{equation}
where $(\sigma_g(\vb*{n}))_i = n_{\sigma_g(i)}$ for $i=0,\cdots,m-1$.

(ii) Letting $\ket*{\Psi_\text{in}}$ and $\ket*{\Psi_\text{out}}$ be the input and output states of an LOC $U$, respectively,
\begin{equation}\label{evolution_formula}
    \ket*{\Psi_\text{in}} \in V^{G}_{\lambda} \Leftrightarrow \ket*{\Psi_\text{out}} \in V^{G'}_{\lambda'},
\end{equation}
where $G'=UGU^\dag \subset \mathcal{M}_m$ and $\lambda'(UgU^\dag)=\lambda(g)$ for $g \in G$.

(iii) For $\ket*{\Psi}\in V^G_\lambda$ and $\ket*{\vb*{n}}$,
\begin{equation}\label{destructive_formula}
    \abs{\bra{\vb*{n}}\ket*{\Psi}}=0
\end{equation}
when $\rho_\lambda^G \ket*{\vb*{n}}=0$ holds, and 
\begin{equation}\label{orbit_formula}
    \abs{\bra{\vb*{n'}}\ket*{\Psi}}=\text{const.}
\end{equation}
for any Fock states $\ket*{\vb*{n'}} \in \mathcal{O}_{\vb*{n}} = \{\ket*{\sigma_g(\vb*{n})}|g \in G\}$.
\end{theorem}

Figure~\ref{image_BSF} is a conceptual diagram of BSF, where the state evolutions are expressed as the stabilizer change from Eq.~\eqref{evolution_formula}.
The dimensions of stabilizer spaces increase with $n$, and the stabilizer state is not uniquely determined for multi-photon states.
Thus, For a given stabilizer group and state, determining whether the state is included in the stabilizer space is meaningful; this can be achieved using Eq.~\eqref{projection_formula}.
From Eq.~\eqref{destructive_formula}, the part of measurement outcomes does not occur; this is a consequence of stabilizer spaces with different eigenvalues orthogonal.
In addition, from Eq.~\eqref{orbit_formula}, a set of measurement outcomes occurs with equal probability~\footnote{Eq.~\eqref{orbit_formula} is insufficient to fully determine the measurement probabilities in general. By contrast, the QSF counterpart of Eq.~\eqref{orbit_formula} shows that all possible outcomes occur with equal probability, yielding the measurement probabilities themselves.}.

If the output-state stabilizer group consists of diagonal matrices, the situation becomes simpler; then, Cor.~\ref{interference} indicates the condition of measurement outcomes for destructive interferences to occur, unifying the destructive interferences in various LOCs.
\begin{corollary}\label{interference}
Let an Abelian group $G\subset \mathcal{M}_m$, an LOC $U$ that simultaneously diagonalizes $G$ to $G'=UGU^\dag$, and an input state $\ket*{\Psi_\text{in}} \in V^{G}_{\lambda}$.
Then, the measurement outcomes $\ket*{\vb*{n}}$ satisfying
\begin{equation}\label{condition}
    \frac{1}{\abs{G}} \sum_{g\in G'} \lambda'(g)^{-1} \prod_{i=0}^{m-1} [B(g')]_{ii}^{n_i} = 0
\end{equation}
are not obtained; i.e.,
\begin{equation}
    \abs{\bra{\vb*{n}}\ket*{\Psi_\text{out}}}=0
\end{equation}
for the output state $\ket*{\Psi_\text{out}}$.
\end{corollary}

To derive nontrivial results based on Thm.~\ref{main_thm}, we need to choose an input-state stabilizer group $G \subset \mathcal{M}_m$ and an LOC $U \notin \mathcal{M}_m$ such that the output-stabilizer group is $UG U^\dag \subset \mathcal{M}_m$.
The Pauli group $\mathcal{P}$ is a subgroup of $\mathcal{M}_m$.
Thus, letting $G\subset \mathcal{P}$ and $U \in \mathcal{C}$, where $\mathcal{C}$ is the Clifford group, the output-state stabilizer group is a subgroup of $\mathcal{M}_m$ as $UGU^\dag \subset \mathcal{P} \subset \mathcal{M}_m$.
Therefore, various results in QSF are expected to be transferable to BSF.
However, unclear correspondences between them remain, such as those for partial measurements.
The Pauli group and QSF can be generalized to composite systems of qudits with different dimensions~\cite{Van_den_Nest2012-oh,Bermejo-Vega2012-he}, and its Clifford group can be decomposed to tensor products of Fourier matrices and the other matrices, which are monomial~\cite{Tolar2018-bf}.
Thus, only tensor products of Fourier matrices need to be considered as transfer matrices, provided the effect of the part represented by monomial matrices can be neglected.
We represent the $d$-dimensional identity, Pauli-X, Pauli-Z, and discrete Fourier matrices by $I_d$, $X_d$, $Z_d$, and $F_d$, respectively, where $F_dX_dF_d^\dag = Z_d$.

Generalizing stabilizer groups in BSF to non-Abelian groups beyond the analogy with QSF is worthwhile because such generalizations give additional information on the states~\footnote{QSFs with non-Abelian stabilizer groups have also been studied~\cite{Ni2015-gh}. In BSF, however, such generalization is necessitated by the $n$-boson representation.}.
Consider an input state $\ket*{\overline{n} \overline{n}}$ for an integer $\overline{n}$ and LOC $F_2$. 
With non-Abelian stabilizer group $G=\langle X_2,Z_2\rangle$, 
$\ket*{\overline{n} \overline{n}}\in V_\lambda^G$, where $\lambda(X_2)=1$ and $\lambda(Z_2)=(-1)^{\overline{n}}$.
Then, the output state is included in $V_{\lambda'}^{G'}$, where $G'=UGU^\dag=G$, $\lambda'(Z_2)=1$, and $\lambda'(X_2)=(-1)^{\overline{n}}$.
Therefore, from Thm.~\ref{main_thm}, each detector only detects even number of photons; furthermore, the outcomes of $\ket*{n_1,n_2}$ and $\ket*{n_2,n_1}$ occur with the same probability for integers $n_1$ and $n_2$.
Further studies using representation theory may be needed for non-Abelian stabilizer groups. 

Next, we consider applications of BSF to LOQC.
A quantum instrument is a set of quantum operations for every measurement outcome~\cite{Watrous2018-la}.
In LOQC, we are interested in what quantum instruments are realizable with general LOCs.
Here, we refer to such realizable quantum instruments as bosonic quantum instruments (BQIs).
Note that various qubit encodings into Fock space exist, and BQIs are defined for each encoding.
Analysis of an LOC as a BQI can be reduced to the problems of discriminating all possible input states.
We approach these problems by considering the stabilizer space to which the measured Fock state belongs rather than the measured Fock state itself.
The information obtained by this approach is limited but, in some cases, enough to discriminate all possible input states.
\begin{definition}
    Let an Abelian group $G\subset \mathcal{M}_m$ and an LOC $U$ that simultaneously diagonalizes $G$ to $G'=UGU^\dag$.
    In BSF, a Fock-basis measurement following $U$ is called the measurement of $G$, and its results are expressed as $g=\lambda(g)$ when a Fock state $\ket*{\vb*{n}} \in V^{G'}_{\lambda'}$ is measured, where $\lambda'(UgU^\dag)=\lambda(g)$ for $g \in G$.
\end{definition}
\begin{corollary}\label{stab_meas_formula}
    For the measurement of an Abelian group $G$ for a state $\ket*{\Psi}$ in BSF, the probability of obtaining $g=\lambda(g)$ for each $g\in G$ is $\norm*{\rho^G_\lambda \ket*{\Psi}}^2$.
\end{corollary}
Cor.~\ref{stab_meas_formula} gives an efficiently computable formula for the measurement probabilities of the stabilizers.
We note that it should only be applied to the part that is not easily computable, as
some output-state information is lost when BSF is applied.
When a transfer matrix is block-diagonal, i.e., an LOC consists of disjoint LOCs, Cor.~\ref{stab_meas_formula} should be applied to each block following division into cases based on the number of photons incident on each block~\footnote{Alternatively, we can introduce diagonal stabilizers that give unique values dependent on the number of photons incident on each block.}.

\begin{figure}[tbp]
    \begin{minipage}[b]{0.45\linewidth}
        \centering
        \subcaption{}
        \includegraphics[scale=0.80, trim=0 0 0 0, clip]{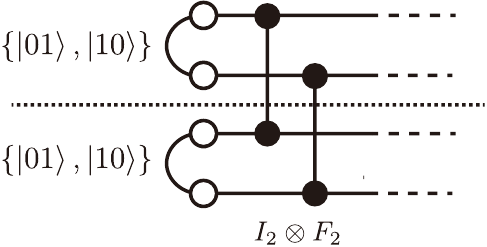}
        \label{EL_scheme}
        \vspace{-15pt}
        \subcaption{}
        \includegraphics[scale=0.80, trim=0 0 0 0, clip]{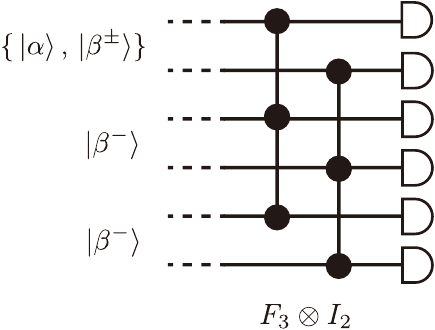}
        \label{F3_I2}
    \end{minipage}
    \begin{minipage}[b]{0.50\linewidth}
        \centering
        \subcaption{}
        \includegraphics[scale=0.70, trim=0 0 0 0, clip]{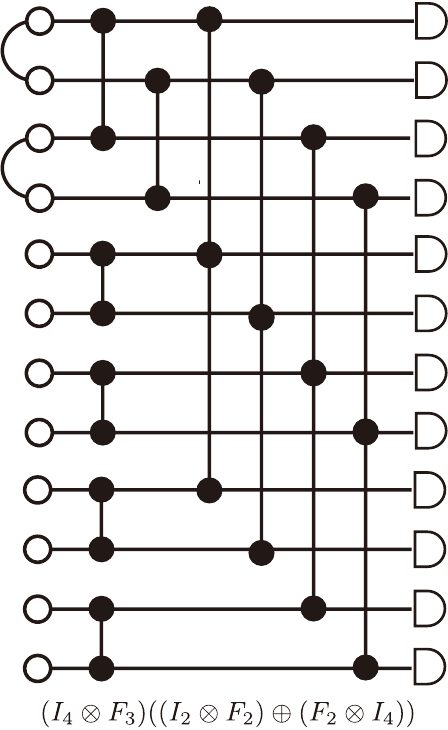}
        \label{whole_LOU}
    \end{minipage}
    \caption{(a) Part of Ewert--van Loock scheme. (b) Discrimination of $\ket*{\alpha}$ and $\ket*{\beta^\pm}$ for $m=3$. (c) Entire LOC for $m=3$.
    The $k$ connected black circles represent LOC $F_k$ for integers $k$. Two connected white circles form a qubit, and an unconnected white circle represents a photon.}
    \label{diagram_LOCs}
\end{figure}

\textit{Bell-state discrimination with single photons.---}
We consider a Bell-state discrimination with ancillary single photons.
The starting point is the Ewert--van Loock scheme, which has 75\% success probability with four single photons~\cite{Ewert2014-uw, Note10}.
\footnotetext[10]{Another possible starting point is the Grice scheme~\cite{Grice2011-hn}, for which the resulting scheme is simpler than that proposed herein; however, Bell states are required as ancillae instead of single photons.}
The Bell states $\ket*{\psi^\pm}=\frac{1}{\sqrt{2}}(\ket*{1001}\pm\ket*{0110})$ and $\ket*{\phi^\pm}=\frac{1}{\sqrt{2}}(\ket*{1010}\pm\ket*{0101})$ in dual-rail encoding are transformed by LOC $I_2 \otimes F_2$ as follows:
\begin{align}
    B(I_2\otimes F_2)\ket*{\psi^+} &=\frac{1}{\sqrt{2}}(\ket*{\alpha}\ket*{00}+\ket*{00}\ket*{\alpha}), \\
    B(I_2\otimes F_2)\ket*{\psi^-} &= \frac{1}{\sqrt{2}}(\ket*{10}\ket*{01}-\ket*{01}\ket*{10}), \\
    B(I_2\otimes F_2)\ket*{\phi^\pm} &= \frac{1}{\sqrt{2}}(\ket*{\beta^\pm}\ket*{00}+\ket*{00}\ket*{\beta^\pm}),
\end{align}
where $\ket*{\alpha}=\ket*{11}$ and $\ket*{\beta^\pm}=\frac{1}{\sqrt{2}}(\ket*{20}\pm\ket*{02})$.
The key technique is to split the four modes into the first and last two modes (Fig.~\ref{EL_scheme}).
Here, $\ket*{\psi^-}$ is distinguished by the presence of one photon in each part.
By preparing the same setup for the first and last two modes, the Bell-state discrimination is reduced to the discriminations of $\ket*{\alpha}$ and $\ket*{\beta^\pm}$.
In the following, we use $\ket*{\beta^-}$ as an ancilla because it is generated from two single photons as $B(F_2)\ket*{\alpha}=\ket*{\beta^-}$.

Consider the discrimination of $\ket*{\alpha}$ and $\ket*{\beta^\pm}$ with the LOC consisting of $I_2 \otimes F_m$, $2m$ detectors, and ancilla photons $\ket*{\beta^-}^{\otimes m-1}$ (Fig.~\ref{F3_I2}).
In BSF, this LOC corresponds to the measurement of $G=\langle Z_2 \otimes I_m, I_2 \otimes X_m\rangle$.
It holds that $Z_2 \otimes I_m = -1$ for $\ket*{\alpha}\ket*{\beta^-}^{\otimes m-1}$ and $Z_2 \otimes I_m = 1$ for $\ket*{\beta^\pm}\ket*{\beta^-}^{\otimes m-1}$.
Thus, $\ket*{\alpha}$ is distinguished.
Furthermore, $I_2 \otimes X_m = 1$ for $\ket*{\beta^-}^{\otimes m}$.
Thus, we conclude that the input state is $\ket*{\beta^+}$ when we obtain the measurement results $Z_2 \otimes I_m = 1$ and $I_2 \otimes X_m \neq 1$.
The measurement probability is calculated from Cor.~\ref{stab_meas_formula} as $1-\norm*{\rho^G_{\lambda=1} \ket*{\beta^+}\ket*{\beta^-}^{\otimes m-1}}=1-m^{-1}$.
Therefore, $\ket*{\beta^+}$ is asymptotically distinguishable, and the average success probability of the Bell-state discrimination converges to $\frac{3}{4}$.
Moreover, as three of four Bell states are almost distinguishable, the failure event almost corresponds to the detection of the other state, $\ket*{\beta^-}$.
Thus, the entire LOC asymptotically corresponds to the Bell measurement.

The actual situation is more complicated because the considered LOC consists of two disjoint LOCs and the states are distinguished according to the number of photons incident on each LOC.
For even $m$, the case with the same number of photons incident on each LOC is critical.
Then, the state corresponding to the input state of $\ket*{\beta^+}$ does not give the measurement result of $I_2 \otimes X_m = 1$, and
$\ket*{\beta^-}$ is distinguished from $\ket*{\beta^+}$.
That is why the average success probability has a maximum value.
By contrast, we expect the average entanglement generated by the LOC to increase with $m$ monotonically.
We quantify this property using the relative entropy of entanglement of quantum measurements~\footnote{The relative entropy of entanglement of quantum states is one of entanglement measures. Here, we consider a similar measure for quantum measurements. See Def.~\ref{measure_def} and Lemma~\ref{RE_lemma} in the Supplemental Material for the definition and related lemma.}.
The entire LOC (Fig.~\ref{whole_LOU}) is fully characterized by identifying the corresponding quantum measurement as a BQI.
By applying the LOC to a set of states and calculating the measurement probabilities based on BSF, we obtain the corresponding BQI as the following Thm.~\ref{bell_thm}.
The success probability and relative entropy of entanglement are calculated from Thm.~\ref{bell_thm} as Cor.~\ref{bell_values}.
\begin{theorem}\label{bell_thm}
Let the LOC consisting of $(I_4 \otimes F_m)((I_2 \otimes F_2) \oplus (F_2 \otimes I_{2(m-1)}))$, $4m$ detectors, and $4(m-1)$ ancillary single photons.
Then, it corresponds to the two-qubit quantum measurement in dual-rail encoding represented by the following Kraus operators: 
\begin{equation}\label{measurment_basis}
    \{\bra{\psi^+}, \bra{\psi^-}, \sqrt{1-m^{-1}}\bra{\phi^+}, K_0, \cdots, K_m \},
\end{equation}
where 
\begin{equation}\label{Kraus_opes}
    K_k = \sqrt{\binom{m}{k}}\frac{1}{2^{m/2}}\qty[\bra{\phi^-} + \frac{(m-2k)}{m}\bra{\phi^+}]
\end{equation}
for $k=0,\dots,m$.
\end{theorem}
\begin{corollary}\label{bell_values}
    For the quantum measurement in Thm.~\ref{bell_thm}, the average success probability of Bell-state discrimination is
    \begin{equation}\label{success_probability}
        P_m = \frac{1}{4}\qty(3-\frac{1}{m}+\binom{m}{m/2}\frac{1}{2^m})
    \end{equation}
    for even $m$, and the relative entropy of entanglement is
    \begin{equation}\label{entanglement_measurement}
        E_m = \frac{1}{4}\qty[3-\frac{1}{m} + \sum_{k=0}^{m} \binom{m}{k}\frac{1}{2^{m-1}}F\qty(\frac{k}{m})],
    \end{equation}
    where $F(x)=\sum_{p=\{x,1-x\}}-p^2\log_2{\frac{p^2}{x^2+(1-x)^2}}$.
    For $m \rightarrow \infty$, it holds that $P_m \rightarrow \frac{3}{4}$ and $E_m \rightarrow 1$.
    For $m=8$, $P_m$ has a maximum value of $\frac{403}{512}$ .
\end{corollary}
\begin{figure}[tbp]
    \centering
    \includegraphics[scale=0.7, trim=0 0 0 0, clip]{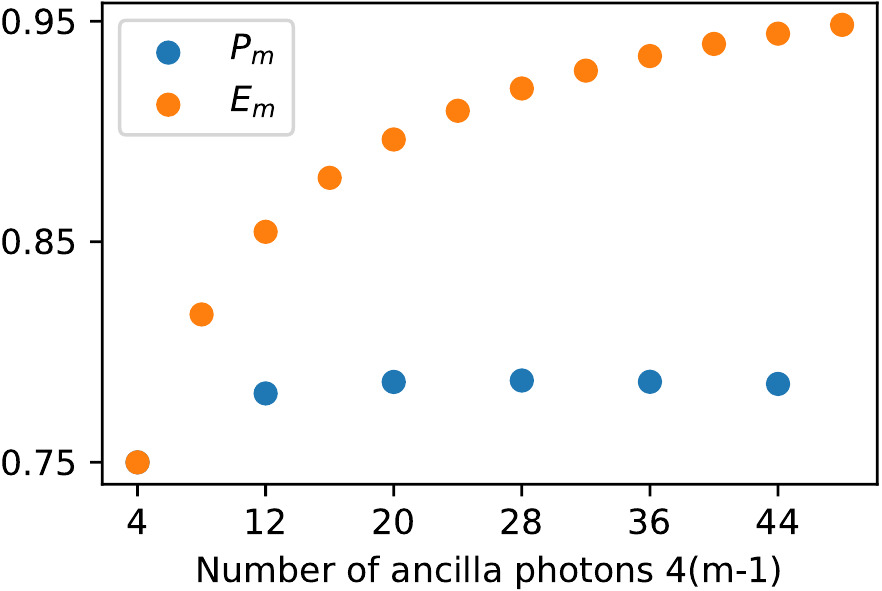}
    \caption{Success probability $P_m$ and relative entropy of entanglement $E_m$ for number of ancilla photons $4(m-1)$ up to $m=12$.}
    \label{plot}
\end{figure}
Figure~\ref{plot} shows $P_m$ and $E_m$ values for small integers $m$.
$P_8\simeq 0.787$ is higher than the previous highest value of $\frac{25}{32} \simeq 0.781$ in Ref.~\cite{Ewert2014-uw}.
Furthermore, $E_m \rightarrow 1$ means that a near-deterministic maximally entangling measurement is possible without ancillary entanglement~\footnote{$E_m \rightarrow 1$ does not mean that the quantum measurement asymptotically approaches the Bell measurement. In this case, however, the corresponding quantum measurement asymptotically approaches the Bell measurement, as implied by Eq.~\eqref{Kraus_opes}}.
$F_{m}$ in this scheme can be replaced with other matrices.
For example, replacing $F_{2^m}$ with $F_2^{\otimes m}$ yields similar results when $X_{2^m}$ is replaced with $X_2\otimes I_{2(m-1)}, \cdots, I_{2(m-1)} \otimes X_2$ in the stabilizer group~\footnote{See Lemma~\ref{probability} in the Supplemental Material for the general requirement for the stabilizer groups.}.
The scheme using $F_m$ is most flexible regarding the LOC size; however, in particular physical systems, implementing the alternative may be easier. 
The key technique of this scheme is to use the interference of input states and multiple ancilla states identical to one of the input states.
It can be applied to other schemes with a fixed size of entangled ancillae.

\textit{Conclusion.---}
We introduced the framework to analyze LOCs named BSF and derived several analytical formulas.
We also proposed a Bell-state discrimination scheme with ancillary single photons and identified the corresponding BQI, based on BSF.
The results obtained by BSF hold for various LOCs, regardless of the number of photons.
Thus, BSF aid the exploration of a wide range of new BQIs.
Moreover, the properties of BSF itself have not been completely explored, especially with regard to their generalization.
Such investigations will reveal further hidden structures of LOCs.

\textit{Note.---}
New suppression laws not covered by Cor.~\ref{interference} have recently been proposed~\cite{Bezerra2023-cm}.

T. Yamazaki thanks Shintaro Minagawa for useful discussions on entanglement measures of quantum measurements.
This work was supported by JST Moonshot R\&D JPMJMS2066, JPMJMS226C, MEXT/JSPS KAKENHI JP20H01839, JP20J20261, JP21H04445, and Asahi Glass Foundation.
\makeatletter\@input{perm_bell_sup_aux.tex}\makeatother
\bibliography{paperpile.bib}
\end{document}

\section{Supplemental material}
\subsection{A. Proof of Theorem~\ref{main_thm}}

\begin{lemma}\label{monomial}
    For $g \in \mathcal{M}_m$,  
    \begin{equation}
        [B_n(g)]_{\vb*{n'}\vb*{n}}=\delta_{\sigma_g(\vb*{n'}),\vb*{n}}  \prod_{i=0}^{m-1} [D(g)]_{ii}^{n_i}.
        \label{monomial_formula}
    \end{equation}
\end{lemma}
\begin{proof}
Because $B_n$ is a group homomorphism, $B_n(g)= B_n(P_{\sigma_g})B_n(D(g))$.
From the definition, 
\begin{equation}\label{permutation_formula}
    B(P_{\sigma_g})\ket*{\vb*{n}}=\prod_{i}\frac{1}{\sqrt{n_i!}}\qty\Big(\sum_{j}(a^\dag_j [P_{\sigma_g}]_{ji}))^{n_i}\ket*{\vb*{0}} 
    =\prod_{i}\frac{1}{\sqrt{n_i!}}(a^\dag_{\sigma_g(i)})^{n_i}\ket*{\vb*{0}} = \ket*{{\sigma_g}^{-1}(\vb*{n})}
\end{equation}
and
\begin{equation}\label{diagonal_formula}
    B(D(g))\ket*{\vb*{n}}=\prod_{i}\frac{1}{\sqrt{n_i!}}\qty\Big(\sum_{j}(a^\dag_j [D(g)]_{ji}))^{n_i}\ket*{\vb*{0}} 
    =\prod_{i}[D(g)]_{ii}^{n_i}\ket*{\vb*{n}}.
\end{equation}
Therefore, we have Eq.~\eqref{monomial_formula}.
\end{proof}

\begin{lemma}\label{projector}
Let 
\begin{equation}
    \rho^G_\lambda = \frac{1}{\abs{G}}\sum_{g\in G} \lambda(g)^{-1} B(g)
\end{equation}
for an Abelian group $G$. Then, $\rho^G_\lambda$ is the projector onto $V^G_\lambda$.
\end{lemma}
\begin{proof}
For any state $\ket*{\Psi} \in V^G_\lambda$, $\rho^G_\lambda \ket*{\Psi}=\ket*{\Psi}$ holds.
It also holds that 
\begin{align}
    B(g) \rho^G_\lambda &= \frac{1}{\abs{G}}\sum_{h\in G} \lambda(h)^{-1} B(gh) 
    =\frac{1}{\abs{G}}\sum_{h\in G} \lambda(g^{-1}h)^{-1} B(h) =
    \lambda(g)\rho^G_\lambda 
\end{align}
for any $g \in G$.
Therefore, $\rho^G_\lambda \ket*{\Psi} \in  V^G_\lambda$ for any state $\ket*{\Psi}$, and $(\rho^G_\lambda)^2 =\rho^G_\lambda$ .
\end{proof}
Lemma~\ref{projector} can be regarded as a special case of the projection formula in representation theory~\cite{Fulton1991-gz}, where the considered group is Abelian.
\begin{proof}[Proof of Theorem~\ref{main_thm}]
(i) We have
\begin{align}
    \norm{\rho_\lambda^G \ket*{\Psi}}^2 &= \sum_{\vb*{n},\vb*{n'}}  c_{\vb*{n'}}^*  c_{\vb*{n}} \bra{\vb*{n'}} \rho_\lambda^G \ket*{\vb*{n}}
    = \sum_{\vb*{n},\vb*{n'}}  c_{\vb*{n'}}^*  c_{\vb*{n}} \frac{1}{\abs{G}}\sum_{g\in G}\lambda(g)^{-1} \bra{\vb*{n'}} B(g) \ket*{\vb*{n}} \notag \\
    &= \frac{1}{\abs{G}}\sum_{g\in G} \lambda(g)^{-1}\sum_{\vb*{n}}c_{\sigma^{-1}_g(\vb*{n})}^* c_{\vb*{n}}\prod_{i=0}^{m-1} D(g)_{i,i}^{n_i},
\end{align}
where the second and third equalities hold from Lemmas~\ref{projector} and~\ref{monomial}, respectively.

(ii) If $\ket*{\Psi_\text{in}}\in V_\lambda^G$ holds, $\ket*{\Psi_\text{out}}=B(U)\ket*{\Psi_\text{in}}$ satisfies 
\begin{equation}
    B(UgU^\dag)B(U)\ket*{\Psi_\text{in}}=B(U)B(g)\ket*{\Psi_\text{in}}=\lambda(g)B(U)\ket*{\Psi_\text{in}}
\end{equation}
for any $g\in G$. 
Thus, letting $G'=UGU^\dag$ and $\lambda'(UgU^\dag)=\lambda(g)$ for $g \in G$, it holds that  $\ket*{\Psi_\text{out}} \in V^{G'}_{\lambda'}$. 
The converse also holds.

(iii) $\bra{\vb*{n}}\ket*{\Psi}=\bra{\vb*{n}}\rho_\lambda^G \ket*{\Psi}=0$ if $\rho_\lambda^G\ket*{\vb*{n}}=0$. 
For any $g\in G$, 
\begin{equation}
    \lambda(g)\bra{\vb*{n}}\ket*{\Psi}=\bra{\vb*{n}}B(g)\ket*{\Psi}=B(D(g))_{\vb*{n},\vb*{n}}\bra{\sigma_g(\vb*{n})}\ket*{\Psi},
\end{equation} 
where Lemma~\ref{monomial} is applied.
Thus, $\abs{\bra{\vb*{n'}}\ket*{\Psi}}=\text{const.}$ for any $\ket*{\vb*{n'}}\in \mathcal{O}_{\vb*{n}}$.
\end{proof}
\begin{proof}[Proof of Corollary~\ref{interference}]
Because $\ket*{\Psi_\text{out}} \in V^{G'}_{\lambda'}$ from Thm.~\ref{main_thm}(ii), Thm.~\ref{main_thm}(iii) yields $\bra{\vb*{n}}\ket*{\Psi_\text{out}}=0$ when $\rho_\lambda^G\ket*{\vb*{n}}=0$.
This condition is equal to Eq.~\eqref{condition} from Thm.~\ref{main_thm}(i), when $\ket*{\Psi}=\ket*{\vb*{n}}$ and $G$ consists of diagonal matrices.
\end{proof}
\begin{proof}[Proof of Corollary~\ref{stab_meas_formula}]
Note that any Fock state is a joint eigenstate of $B(G')$ when $G'$ consists of diagonal matrices, as seen from Eq.~\eqref{permutation_formula}.
Then, the projector can be expressed as $\rho^{G'}_{\lambda'} = \sum_{\ket*{\vb*{n}}\in V^{G'}_{\lambda'}} \dyad{\vb*{n}}$.
From the definition, the probability of obtaining the measurement results $g=\lambda(g)$ for each $g\in G$ is 
\begin{equation}
    \sum_{\ket*{\vb*{n}}\in V_{\lambda'}^{G'}} \abs{\bra{\vb*{n}} B(U)\ket*{\Psi}}^2,
\end{equation}
which is equal to 
\begin{equation}
    \bra{\Psi} B(U^\dag) \rho^{G'}_{\lambda'} B(U)\ket*{\Psi}=\norm{\rho^{G}_{\lambda} \ket*{\Psi}}^2.
\end{equation}
\end{proof}

\subsection{B. Proof of Theorem~\ref{bell_thm}}
\begin{definition}\label{Dicke}
    The Dicke states are
    \begin{equation}
        \ket*{D^m_{k}} = \binom{m}{k}^{-\frac{1}{2}} \sum_{\substack{x\in \{0,1\}^{m}\\ wt(x)=k}} \ket*{x},
    \end{equation}
    where $wt(x)$ is a Hamming weight of $x$.
\end{definition}
\begin{lemma}\label{probability}
    Assume an Abelian transitive subgroup of symmetry group $\Sigma \subset S_m$, where there exists unique $\sigma \in \Sigma$ such that $\sigma(0)=i$ for each $i\in \{0,\cdots, m-1\}$.
    Letting ${P}_j(\theta)$ be the operator acting on the $j$th modes as ${P}_j(\theta)\ket*{{k}}_j=e^{ik\theta}\ket*{k}_j$, 
    \begin{equation}\label{prob_Dicke_formula}
        \norm*{\rho^{\Sigma}_1 {P}_0(\theta) \ket*{D^m_{k}}}^2 = \abs{\frac{(m-k)+e^{i\theta}k}{m}}^2,
    \end{equation}
    where $\rho^{\Sigma}_1$ is the projector onto the joint eigenstate of $\{B(P_\sigma)|\sigma \in \Sigma\}$ with eigenvalues of $1$.
\end{lemma}
\begin{proof}
From the definition, the Dicke states are invariant under any permutation of the modes; that is, 
\begin{equation}
    B(P_\sigma)\ket*{D^m_{k}}=\ket*{D^m_{k}}
\end{equation}
for any $\sigma \in S_m$.
Further, because ${P}_i(\theta)$ acts on the $i$th modes, 
\begin{equation}
    B(P_\sigma){P}_i(\theta)B(P_\sigma)^\dag={P}_{\sigma(i)}(\theta),
\end{equation}
from Eq.~\eqref{permutation_formula}.
From these equations and Lemma~\ref{projector}, 
\begin{equation}\label{cal1}
    \rho^{\Sigma}_1 {P}_0(\theta) \ket*{D^m_{k}} = \frac{1}{\abs{\Sigma}}\sum_{\sigma\in \Sigma} {P}_{\sigma(0)}(\theta) \ket*{D^m_{k}}.
\end{equation}
The group $\Sigma$ being an Abelian transitive group implies that there is only one element $\sigma \in \Sigma$ such that $\sigma(0)=i$ for each $i\in \{0,\cdots, m-1\}$, and that $\abs{\Sigma}=m$.
Therefore, Eq.~\eqref{cal1} is equivalent to
\begin{equation}
    \frac{1}{m}\sum_{i=0}^m {P}_i(\theta) \ket*{D^m_{k}} = \frac{(m-k)+e^{i\theta}k}{m}\ket*{D^m_{k}}.
\end{equation}
Therefore, Eq.~\eqref{prob_Dicke_formula} holds.
\end{proof}

\begin{proof}[Proof of Theorem~\ref{bell_thm}]
From the main text, the remainder of the proof of Eq.~\eqref{measurment_basis} involves showing that the LOC consisting of $I_2 \otimes F_m$, $2m$ detectors, and ancilla photons $\ket*{\beta^-}^{\otimes m-1}$ corresponds to the measurement on $\ket*{\beta^\pm}$ expressed as
\begin{equation}\label{measurment_basis2}
    \{\sqrt{1-m^{-1}}\bra{\beta^+}, K'_0, \cdots, K'_m \},
\end{equation}
where 
\begin{equation}\label{Kraus'}
    K'_k = \sqrt{\binom{m}{k}}\frac{1}{2^{m/2}}\qty[\bra{\beta^-} + \frac{(m-2k)}{m}\bra{\beta^+}].
\end{equation}
Here, the LOC $I_2 \otimes F_m = F_m \oplus F_m$ preserves the number of photons incident on each block of the LOC. 
Thus, we can distinguish the input state according to these numbers.
For simplicity, we introduce the notation $\ket*{\overline{0}}=\ket*{20}$ and $\ket*{\overline{1}}=-\ket*{02}$.
Let $\overline{P}_i(\theta)$ be the operator acting as $\overline{P}_i(\theta)\ket*{x}_i=e^{ix\theta} \ket*{x}_i$ for $x=\{\overline{0},\overline{1}\}$. 
Then, we can expand the state as
\begin{equation}
    \ket*{\beta^+}\ket*{\beta^-}^{\otimes m-1} = \frac{1}{2^{m/2}}\overline{P}_0(\pi)(\ket*{\overline{0}}+\ket*{\overline{1}})^{\otimes m} 
    =\frac{1}{2^{m/2}} \sum_{k=0}^{m}\binom{m}{k}^{\frac{1}{2}} \overline{P}_0(\pi) \ket*{\overline{D}_k^m},
    \label{expansion}
\end{equation}
where $\ket*{\overline{D}_k^m}$ is the Dicke state in Def.~\ref{Dicke} defined for $\{\ket*{\overline{0}},\ket*{\overline{1}}\}^{\otimes m}$.
The number $k$ in Eq.~\eqref{expansion} is obtained from the measurement result.
The LOC $I \otimes X_m$ corresponds to the permutation of $(0\cdots m-1)$ of the modes of $\{\ket*{\overline{0}},\ket*{\overline{1}}\}^{\otimes m}$.
Thus, Lemma~\ref{probability} yields 
\begin{equation}\label{probability_cal}
    \norm*{\rho^{G}_1 {\overline{P}}_0(\pi) \ket*{\overline{D}^m_{k}}}^2 = \norm*{\rho^{C_n}_1 {P}_0(\pi) \ket*{D^m_{k}}}^2 = \qty\Big(\frac{m-2k}{m})^2,
\end{equation}
where $G=\langle I \otimes X_n\rangle$ and $C_m = \langle (0 \cdots m-1)\rangle$.
Because $\ket*{\beta^-}^{\otimes m} \in V_1^G$, we can distinguish $\ket*{\beta^+}$ from $\ket*{\beta^-}$ when we measure $I \otimes X_n \neq 1$, the total probability of which is
\begin{equation}
    \frac{1}{2^m}\sum_{k=0}^{m}\binom{m}{k}\qty\Big(1-\qty\Big(\frac{m-2k}{m})^2)=1-\frac{1}{m}.
\end{equation}
In contrast, when we obtain $I \otimes X_n = 1$, we can expect the corresponding Kraus operators to be Eq.~\eqref{Kraus'}, from Eq.~\eqref{probability_cal}.
To completely identify the measurement basis, we need to perform a kind of measurement tomography, in which we consider multiple bases as the input states and reconstruct the measurement corresponding to this setup.
For input states
\begin{equation}
    \frac{1}{\sqrt{2}}(\ket*{\beta^-}_0\pm i \ket*{\beta^+}_0) = \frac{e^{\pm i \pi/4}}{\sqrt{2}} P_0(\mp \frac{\pi}{2})_0 (\ket*{\overline{0}}+\ket*{\overline{1}}),
\end{equation}
We can calculate the measurement probabilities from Lemma~\ref{probability} in the same manner.
Therefore, we can ensure that the considered setup corresponds to the measurement represented by Eq.~\eqref{measurment_basis2}.
Thus, the entire setup corresponds to the quantum measurement on two qubits represented by Eq.~\eqref{measurment_basis}.
\end{proof}
Following Ref.~\cite{Kim2022-oj}, we introduce the relative entropy of entanglement of quantum measurements as an entanglement measure. 
Computing this measure is generally difficult; however, we can explicitly calculate it in the special case as follows.
\begin{definition}\label{measure_def}
    For the $d$-dimensional systems $A$ and $B$, consider a quantum measurement on the composite system $AB$, which is described by positive operator-valued measure (POVM)  elements as $M_{AB}=\{M_x\}$.
    The relative entropy of entanglement of a quantum measurement $M_{AB}$ is
    \begin{equation}
        E(M_{AB})=\frac{1}{d^2} \min_{F_{AB} \in \text{sepM(A:B)}}\sum_x S(M_x||F_x),
    \end{equation}
    where $S( \cdot || \cdot )$ is the quantum relative entropy and $\text{sepM(A:B)}$ is the set of separable measurements on $A$ and $B$~\cite{Watrous2018-la}.
\end{definition}
\begin{lemma}\label{RE_lemma}
    Assume that the rank of all POVM elements is $1$ in Def.~\ref{measure_def}. Then, 
    \begin{equation}\label{RE_formula}
        E(M_{AB}) = \frac{1}{d^2} \sum tr(M_x) E_s \qty\Big(\frac{M_x}{\tr(M_x)}),
    \end{equation}
    where $E_s(\cdot)$ is the entanglement entropy of quantum states.
\end{lemma}
\begin{proof}
    Let $M_x =p_x \rho_x$ with $p_x=\tr(M_x)$, and $F_x=q_x \sigma_x$ with $q_x=\tr(F_x)$.
    Then, 
    \begin{align}\label{RE_calculation}
        E(M_{AB}) 
        &= \frac{1}{d^2} \min_{F_{AB} \in \text{sepM(A:B)}}\sum_x \qty\Big{p_x S(\rho_x||\sigma_x) + p_x \log{\frac{p_x}{q_x}}} \notag \\
        &=\frac{1}{d^2} \min_{F_{AB} \in \text{sepM(A:B)}} \qty\Big{\sum_x p_x S(\rho_x||\sigma_x) + S(p||q)},
    \end{align}
    where $S(p||q)$ is the classical relative entropy of $p=\{p_x\}$ with respect to $q=\{q_x\}$.
    $S(p||q)$ takes the minimum value of $0$ when $q=p$. 
Here, $\rho_x$ is a pure state, and $S(\rho_x||\sigma_x)$ takes the minimum value of $E_s(\rho_x)$~\cite{Vedral1998-fg}.
    Letting $\mathcal{I}$ and $\Delta$ be the identity and completely dephasing channel for any fixed basis, respectively, the state at the minimum is expressed as $\sigma_x=(\mathcal{I} \otimes \Delta)(\rho_x)$.
    Letting $F_{AB}=\{F_x\}=\{p_i (\mathcal{I} \otimes \Delta)(\rho_x)\}_x$, $F_{AB}$ satisfies the conditions for being a quantum measurement, especially as
    \begin{equation}
        \sum_x F_x = \sum_x p_x (\mathcal{I} \otimes \Delta)(\rho_x) = (\mathcal{I} \otimes \Delta)(I) = I,
    \end{equation}
    from $\sum_x p_x \rho_x = \sum_x M_x = I$.
    Thus, the quantum measurement $F_{AB}$ simultaneously minimizes each term in Eq.~\eqref{RE_calculation} and, hence, Eq.~\eqref{RE_formula} holds.
\end{proof}

We use the following Lemma~\ref{known_thm} without proof.
\begin{lemma}\label{known_thm}
    For the binary entropy function $H_b$, 
    \begin{equation}\label{inquality1}
        4p(1-p) \leq H_b(p) \leq (4p(1-p))^{1/ln4}
    \end{equation}
    for $0 \leq p \leq 1$~\cite{Topsoe2001-wd}, and 
    \begin{equation}\label{inquality2}
        \sqrt{\frac{n}{8 k (n-k)}}  < \binom{n}{k}2^{-nH_b(k/n)} \leq  \sqrt{\frac{n}{\pi k (n-k)}} 
    \end{equation}
    for integers $n$ and $1 \leq  k \leq n-1$~\cite{Cover2006-eh}.
\end{lemma}
\begin{lemma}\label{F_function_lemma}
    Let 
    \begin{equation}
        F(x)=-x^2\log_2{\frac{x^2}{x^2+(1-x)^2}}-(1-x)^2\log_2{\frac{(1-x)^2}{x^2+(1-x)^2}}
    \end{equation}
    and 
    \begin{equation}\label{f_function_def}
        f(x)=\frac{4x^2(1-x)^2}{x^2+(1-x)^2}.
    \end{equation}
    Then,  
    \begin{equation}\label{inequality2}
        f(x) \leq F(x) \leq \frac{1}{2}
    \end{equation}
    for $0\leq x \leq 1$ and $f'(x) \geq 0$ for $0\leq x \leq \frac{1}{2}$~\footnote{We could not find the proof of $F'(x) \geq 0$. Therefore, we introduced the function $f$ and proved $f'(x) \geq 0$ instead.}.
\end{lemma}
\begin{proof}
With the binary entropy function $H_b(x)$, $F$ can be expressed as
\begin{equation}
    F(x)=(x^2+(1-x)^2)H_b\qty\Big(\frac{x^2}{x^2+(1-x)^2}).
\end{equation}
For $0\leq x \leq 1$, it holds that  $(4x(1-x))^{1/ln4}\leq 2\sqrt{x(1-x)}$.
Thus, Eq.~\eqref{inquality1} in Lemma~\ref{known_thm} yields the following inequality:
\begin{equation}\label{F_ub_formula}
    f(x) \leq F(x) \leq 2x(1-x) \leq \frac{1}{2}.
\end{equation}
The first derivative of $f$ is
\begin{equation}
    f'(x)=\frac{8x(1-x)((1-x)-x)(1-x(1-x))}{(x^2+(1-x)^2)^2} \geq 0
\end{equation}
for $0\leq x \leq \frac{1}{2}$.
\end{proof}
\begin{proof}[Proof of Corollary~\ref{bell_values}]
The average success probability $P_m$  of Bell-state discrimination is calculated as the probability that the maximally mixed input state is projected to one of the Bell states.
For even $m$, the case of $k=\frac{m}{2}$ is special because the corresponding Kraus operator becomes the projection onto a Bell state $\ket*{\phi^-}$, with probability $\binom{m}{m/2}2^{-m}$.
Therefore, $P_m$ is calculated as in Eq.~\eqref{success_probability} in total.
The relative entropy of entanglement $E_m$ can be calculated as in Eq.~\eqref{entanglement_measurement} using Lemma~\ref{RE_lemma}.

From Eq.~\eqref{inquality1} in Lemma~\ref{known_thm}, $P_m$ are upper bounded as
\begin{equation}
    P_m \leq P_u(m) = \frac{1}{4}\qty(3-\frac{1}{m}+\sqrt{\frac{4}{\pi m}}).
\end{equation}
From the calculation, $P_u'(x)<0$ for $x>\pi$. 
Numerically, we confirmed that $P_u(44) \simeq 0.7868$ and that the maximum value of $P_m$ for $m<44$ is 
\begin{equation}
    P_8 = \frac{403}{512} \simeq 0.7871.
\end{equation}
Thus, $P_8$ is the maximum value of $P_m$, from $P_8>P_u(44)$.
Moreover, $P_m \rightarrow \frac{3}{4}$ for $m \rightarrow \infty$ are proved by lower-bounding $P_m$ in the same manner.

Letting 
\begin{equation}
    S(m) = \frac{1}{2^{m-1}}\sum_{k=0}^m \binom{m}{k}F\qty\Big(\frac{k}{m}),
\end{equation}
$S(m)\rightarrow 1$ yields $E_m \rightarrow 1$ for $m\rightarrow \infty$. 
Because $F(x)\leq \frac{1}{2}$ for $0\leq x \leq 1$, from Lemma~\ref{F_function_lemma}, it holds that
\begin{equation}
    S(m) \leq \frac{1}{2^m}\sum_{k=0}^m \binom{m}{k} = 1.
\end{equation}
Introducing a real number $0< c < \frac{1}{2}$, we obtain the lower bound of $S_m$ as
\begin{equation}
    S _m > \frac{1}{2^{m-1}} \sum_{k=\lceil cm \rceil}^{m-\lceil cm \rceil} \binom{m}{k}F\qty\Big(\frac{k}{m}) \\
    > f(c) \frac{1}{2^{m-1}} \sum_{k=\lceil cm \rceil}^{m-\lceil cm \rceil} \binom{m}{k},
\end{equation}
where $f$ is defined in Eq.~\eqref{f_function_def}, and the second inequality holds from $F(x) \geq f(x)$ and $f'(x) \geq 0$ for $0\leq x< \frac{1}{2}$ in Lemma~\ref{F_function_lemma}.
Further, Eq.~\eqref{inquality1} in Lemma~\ref{known_thm} implies the inequality 
\begin{equation}
    \binom{m}{k} \leq 2^{m H_b(\frac{k}{m})}
\end{equation}
for $0\leq k \leq n$. 
Thus, 
\begin{equation}
    \sum_{k=0}^{\lceil cm \rceil-1} \binom{m}{k}  \leq c m 2^{m H_b(c)}.
\end{equation}
Finally, we have the following lower bound: 
\begin{equation}\label{lower_bound}
    S(m) > 2f(c)(1-2cm 2^{-m(1-H_b(c))}).
\end{equation}
It holds that  $H_b(c)<1$ for $c\neq \frac{1}{2}$ and $f(\frac{1}{2})=\frac{1}{2}$.  
Thus, Eq.~\eqref{lower_bound} implies that $S(m)\rightarrow 1$ for $m\rightarrow \infty$ and $c\rightarrow \frac{1}{2}$.
\end{proof}
\bibliography{paperpile.bib}
\makeatletter\@input{perm_bell_aux.tex}\makeatother